# Age-sensitive bibliographic coupling with an application in the history of science

Sándor Soós
Dept. of Science Policy and Scientometrics, Library of the Hungarian Academy of Sciences
soossand@caesar.elte.hu

**Abstract.** In science mapping, bibliographic coupling (BC) has been a standard tool for discovering the cognitive structure of research areas, such as constituent subareas, directions, schools of thought, or paradigms. Modelled as a set of documents, research areas are often sorted into document clusters via BC representing a thematic unit each. In this paper we propose an alternative method called age-sensitive bibliographic coupling: the aim is to enable the standard method to produce historically valid thematic units, that is, to yield document clusters that represent the historical development of the thematic structure of the subject as well. As such, the method is expected to be especially beneficial for investigations on science dynamics and the history of science. We apply the method within a bibliometric study in the modern history of bioscience, addressing the development of a complex, interdisciplinary discourse called *the Species Problem*. As a result, a quantitative and qualitative comparison of the standard and the proposed method of bibliographic coupling will be reported, together with a pilot study on the cognitive–historical structure of the Species Problem, regarding an important fragment of the discourse.

**Keywords:** bibliographic coupling, science mapping, history of science, science dynamics, species problem, citation analysis

**Introduction**

Bibliographic coupling (BC) is a long-established method in science mapping. Its main aim is to detect, within a set of publications, groups or clusters that share a common intellectual background, and, therefore, can be conceived as each representing a particular research problem, program, approach or school, depending on the interpretation. To this effect, the method relies on references, usually conceptualized as conveying the intellectual background of the corresponding papers. The basic principle is that the relatedness of any two papers is a function of the number of references they have in common.

Since the introduction of the method within bibliometrics (Kessler 1963), the method of BC has been effectively applied in many contexts, basically in its original form. In this paper we propose a refinement of BC that takes into account a further parameter of common references: beyond their (usually normalized) number it also incorporates the (respective) age of them. We call this method *age-sensitive bibliographic coupling*. The reason for and our expectations on this alternative method is best communicated with the help of an analogy from biological systematics.

A striking similarity between reference-based science mapping and evolutionary biosystematics is that both attempts to detect groups of related actors based on common ancestors. In the case of science mapping, biological descendancy is to be replaced by citation links, or „intellectual descendancy": a reference can be viewed as an ancestor of the citing document. However, as a disanalogy, biosystematics defines the degree of relatedness as conditional on the „age" of common ancestors: on the evolutionary timescale, the more ancient their common ancestor is, the less related two species are, while the more recently they originated from a common predecessor, the closer they stand in systematics. As a result, biosystematics is capable of setting up a categorization where groups also reflect the history of their formation.

We claim that these considerations can be adopted for bibliographic coupling as well to gain similar advantages. Our modified basic principle of BC, therefore, would formulate in the following way: the more recent references any two papers have in common, the higher the degree of their relatedness is. That is, the (intellectual or cognitive) relatedness of any two papers is a function of the (1) number and the (2) age of references they have in common.

Addressing the age of references in bibliometrics is, by far, not a new idea,—consider, for example, the classical Price index (Price 1970), conveying the age distribution of the intellectual background—nor is the assumption that the subset of references published more recently is indicative of the particular direction of research a paper belongs to, as contrasted to „older" references, characterizing the broader thematic context. However, approaches linking these observations to bibliographic coupling have been rather rare. One such example is the study of (van Raan 2005), addressing the behavior of BC. For a sample of documents to be structured by the method, Raan partitioned the set of aggregated references into two age groups based on two consecutive time windows, producing a cohort of „old" references and another of „young" references. The application of BC on sample documents using the old cohort and the young cohort, respectively, resulted in similarity networks within the sample with different structural characteristics (degree distribution). Based on these results, Raan argued that the young cohort, that is, recent references, is better suited to classify documents according to their intellectual relatedness, which is in accord with the assuption on the role of immediate cognitive ancestors.

As contrasted to this latter approach, our goal is not to filter the set of references so that an improved precision of clustering could be achieved via BC, reflecting exclusively the closest and most timely relations. Instead, we aim at the „whole picture", inside which all relations are made visible, but still (historically) distinguishable: relying on the entire, unfiltered (and aggreageted) list of referred works, we intend to incorporate age as a factor into the method, and potentially obtain clusters being differentiated in this respect: some reflecting a closer, some looser internal historical relatedness. The rationale behind is the same as in the case of biosystematics: by age-sensitive bibliographic coupling we expect to map a research area not only in terms of „thematic directions", but by revealing real, hitorically (causally) connected parts of the discourse. That is, by finding groups reflecting the history of the problem, we aim contribute to the toolkit of the history of science.

**Materials: a corpus on the Species Problem**

In order to test and demonstrate the capacity of the proposed method, we applied it in an attempt to reconstruct the historical development of a rather complex discourse in biology, usually referred to as the *Species Problem*. The Species Problem can be briefly described as a historical debate on what biological species are, and as the related quest for the appropriate definition of species, or species concept for biology. With a long prehistory, dated as back as to Aristotle and Plato, including Darwin's paradigm-shifting work on the nature of species in the XIX. century (milestone #1), the debate expanded in the early XX. century, mainly due to the rediscovery of Darwin's work, and having it integrated with the early (Mendelian) genetics of the era. The new paradigm has been called the *Evolutionary Synthesis* (milestone #2). Since the Synthesis, a plethora of theories has emerged on species, resulting in a variety of competing species concepts. According to a comprehensive review of Mayden (1997), no less than 22 species concept (definitions) exhibit themselves in the contemporary literature of the subject.

Given its complexities, the Species Problem was an ideal candidate for a bibliometric analysis with the proposed method of *asBC* (age-sensitive bibliometric coupling):

(1) The roots of the discourse are centuries-old, while there are several contemporary directions of the debate (and of research) as well. Therefore, the capacity of *asBC* to differentiate between more classical and more recent thematic developments could be tested well.
(2) During its modern history (in the XX. century), many schools of biosystematics contributed to, and competed over the problem, involving—from a data-mining perspective—different topics: theoretical papers as well as empirical ones, the latter focusing on particular subjects of taxonomy (description of taxa). It was of outstanding interest whether the *asBC* was capable of identifying these schools as being pairwise different but internally coherent lines of research.
(3) A nonstandard feature of the Species Problem is its complexity in terms of the contributing scholarly fields, or even disciplines. As we shall see, for example, a proper interaction of evolutionary systematics, on one side, and the philosophy of science (of biology), on the other side, had a significant effect on the present state of the debate. Due to this interdisciplinarity, it is not an easy task to obtain the cognitive structure of the discourse for the historian of science. However, it is a good challenge to the proposed method of science mapping, that may, ideally, help the historian in achieving her goal.

To cover a representative corpus of the modern history of the discourse, bibliographic data were harvested from three databases of the Web of Science, namely, the SCI, the SSCI and the A&HCI (*Science Citation Index*, *Social Science Citation Index* and the *Arts&Humanities Citation Index*, respectively). Also in a attempt to avoid the potential exclusion of relevant works from the corpus, data retrieval was based on a topic-related query, that did not put any constraints on the set of fields, journals, authors etc. entering the sample. The query was defined to include all records related (topicwise) to any of the following terms: „species problem", „species definition", „species concept".

The resulting corpus included N=1605 documents for the period 1975–2011. Since (1) we were primarily interested in the period where the debate became most intense and accelerated (so that empirical methods are helpful to clarify its structure), and (2) the selected data was also required to „contain enough references", potentially reaching back to all historical layers of the debate, we took a smaller time window for our analysis. As confirmed by the distribution of the corpus over publication years (Fig 1), a period starting from the '90s was meeting the intensity criterion, and was late enough to reflect existing directions. We took a fraction of the whole corpus accordingly, covering a decade being a „burst" in the dispute. This fraction contained about 400 records. We pruned it by eliminating those few that did not share any references with the rest (not being related to the problem, in this sense). Our final sample, therefore, contained the fragment of the base corpus published between 1990–2000, with N=386 papers.

**Fig. 1.** *Distribution of the corpus collected on the Species Problem over publication years.*

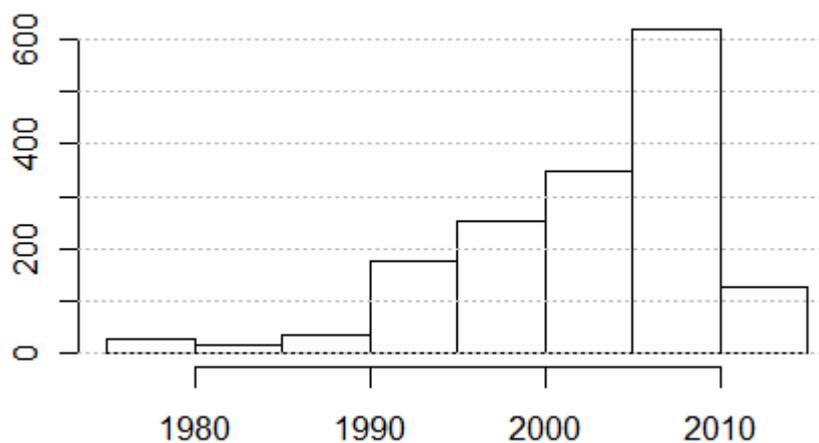

The rest of the paper is organized as follows:

(1) in the next section we define the age-sensitive method of bibliographic coupling. The subsequent section will report the results of analysing our sample via the proposed method in two respects:
(2) first, a quantitative comparison of the results achived by the classical method and the age-sensitive method will be outlined, then
(3) a qualitative comparison of the thematic structure vs. the intellectual history of the species problem, obtained by the classical and the new method, respectively, will be described. It shows, on one hand, how the age-sensitive method changes the cognitive structure revealed by the classical method, and, on the other hand, provides a historical mapping of the species problem.

# The altered method of bibliographic coupling

Bibliographic coupling of a set of publications, in the classical case, is based on the number of their common references, in a pairwise manner. This relation can be conceived as some sort of similarity (or distance) between the two vectors of references of any publications *p1* and *p2*, respectively. These vectors are usually represented as dichotom sequences with values {1,0}, denoting the presence/absence of a publication in the reference set (this method also implies that such vectors are built over the aggregation of the references of each pub in question to make them comparable — practically, these are the rows of a publication-reference incidence matrix).

More formally, the method of determining the relatedness of pubs within classical bibliographic coupling may be presented as follows. Given any two publications $P_1$ and $P_2$, consider the vectors $REF_1$ and $REF_2$ of their respective sets of references. These vectors are best conceived as of length *n*, where *n* is the number of all references belonging to either $P_1$ or $P_2$. Based on these same tuple of referred publications, $REF_1(i)$ denotes whether the *i*-th reference is present among the references of $P_1$, and may take the corresponding value of 1 or 0 (the same goes for $P_2$). In this setting, the basic similarity between the two publications is given by

$$S_{BC}(P_1, P_2) := \sum_{i=1}^{n} REF_1(i) \times REF_2(i).$$

In verbal terms, $S_{BC}(P_1, P_2)$ is the absolute number of references shared among $P_1$ and $P_2$. This amount is usually subject to a normalization procedure accounting for the size of the reference sets of $P_1$ and of $P_2$, respectively, for it is often argued that having the same amount in common out of an extensive background (of which the shared part is a relatively small fraction) makes pubs less related, than if this same amount is a substantial part of the references for any member of the pair. In our study, however, we used this measure in its raw, non-normalized version, mainly for the reasons of comparison with our age-sensitive indicator (see below).

In order to implement the idea of age-sensitive bibliometric coupling, we altered the abovedescribed method of BC in two steps. The procedure was based on the publication–reference incidence matrix constructed from publications in our material.

*Step 1: Weighting*

At first, an indicator of the age of references has been introduced. To systematically account for this feature of reference publications, each component of the presence/absence vectors was weighted according to the publication year of the corresponding reference. This procedure yielded a weighted reference vector for each source publication:

$$REF^W(i) := REF(i) \times f(Pubyear(i)),$$

whereby $REF^W(i)$ is the weighted value of the *i*-th reference within the vector of references *REF(i)*, and this weight is given by a function of the publication year of the *i*-th reference, i.e. *Pubyear(i)*.

Practically, this kind of modification of a presence/absence vector replaces the value „1" of each reference of the source publication with a time-dependent weight, determined by the weighting scheme. In order to reflect our „phylogenetic" notion of relatedness, we defined the particular weighting scheme (the function *f(Pubyear(i))* in the formula) according to the following criteria:

(1) The more recent a shared reference is, the closer relatedness of source documents it should represent.
(2) Classical topic-related literature should reflect distant kinship when referred by pubs, while shared recent literature reflect close kinship. Furthermore, as we intend to amplify the effect of having classical vs. recent common ancestors in drawing relatedness (so that recent kinship and more ancient kinship could be separated), it is assumed that differences between the age of classical (old) publications contribute less to relatedness, than age differences in the recent literature.

In the scheme chosen for weighting, criterion (1) is realized by weights being defined as increasing by publication years. This procedure assures that, when subjected to the similarity measure introduced below, recent references contribute more to document similarity than older ones. Criterion (2) is met by rewarding a reference for being timely, via determining weights as a non-linear function of time (publication years). In particular, we used an exponential function of the rescaled years of publication, the parameters of which were experimentally set to enable the scheme conveying the age effect of the intellectual background, in the case of the topic under study:

$$w(Pubyear) := \text{scale}_1\left(30^{\text{scale}_2(Pubyear)}\right),$$

whereby *Pubyear* is a year of publication (age), *w(.)* is the associated weight, and $\text{scale}_2(Pubyear)$ designates a linear rescaling of the series of publication years within the interval [1,10]. The immediate result was also rescaled within the interval [1,100], indicated by $\text{scale}_1(.)$, to produce intuitive weighting scores for references. Fig 2 graphs the weights associated with years of publication. It can be observed that (due to the distant origins mentioned above) references to the ancient—e.g. medieval or XIX. century—history of the problem, ranging from the XVI. century to the beginning of the XX. century are almost equally weighted, their contibution being kept at a low level. The weighting is becoming rather progressive from the 1960s, and the slope of the curve increases by roughly twenty years (at the beginning of the '80s, and that of the second millenium). This scheme is in accord with our aim to detect the accelerated development of the topic in the XX. century, and also with descriptive studies characterizing similar periods of problem development along the timescale.

**Fig. 2.** *Weights associated with publication years according to the weighting scheme used for the age-based ranking of references*

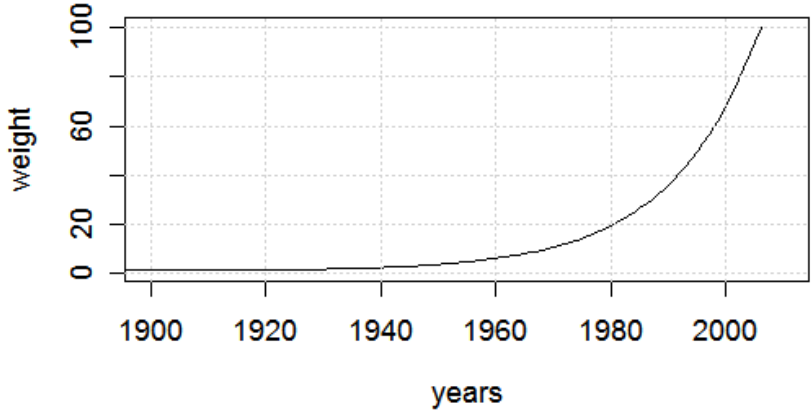

*Step 2: Similarity measure*

As the second step of the method, the degree of relatedness of source documents was calculated based on their weighted reference vector. In particular, we applied the basic similarity measure $S_{BC}$ of bibliographic coupling explicated above, to each pair of such vectors obtained for source documents in the sample. This resulted in a measure

$$S_{BC}^{w}(P_1, P_2) := \sum_{i=1}^{n} REF_1^{w}(i) \times REF_2^{w}(i),$$

Where $S_{BC}^{w}(P_1, P_2)$ is the weighted (or age-sensitive) similarity of publications $P_1$ and $P_2$, while $REF_1^{w}$ and $REF_2^{w}$ are the weighted reference vectors belonging to $P_1$ and $P_2$ respectively.

In practice, according to this measure, the more recent references are being shared by any two publications, the more closely related (similar) these pubs will be. Defined in such a way, this indicator is not normalized (e.g. doesn't control for the number of references that the two publications contain, separatelly), but since we are interested in the effect of age (weighting) of shared references, as disentangled from any other effect, we used the measure as such: this choice allowed us to contrast the results directly with the core of the classical (unnormalized) approach, whereby the same common references are counted but not age-weighted. More importantly, by this definition we obtain a fine-grained relation between publications, even analytically. Consider a publication *P1* that has the same number of common references both with *P2* and *P3*, but with recent publications shared with the former, and with old publications shared with the latter. On the classical account, *P1* is equally similar to *P2* and *P3* (since only the amount of shared references matters). However, on the present account, *P1* is much more similar to *P2* than to *P3*, due to the contribution of recent background literature to the similarity value.

*Clustering of source publications*

Though not specific to the altered procedure of bibliographic coupling discussed so far, a still relevant step of the method is the actual „coupling" (or grouping) of publications, that is, the clustering based on the weighted similarity matrix. For this purpose, a type of hierarchical clustering was selected, and imposed on the distance matrix obtained from the original similarity matrix. We applied the average clustering method, as the resulting hierarchy turned out to be, among those produced by other available methods, best fitted to document distances. (This latter fit was measured by the so-called cophenetic correlation, and yielded a value *cpc* = 0.7)

In order to detect the cluster structure at a fine-grained level, we avoided to cut this cluster tree at a predefined height, as such a trade-off would have resulted in overlooking groups with varying „internal cohesion". Instead, an approach called *dynamic cutting* was utilized, as developed and detailed in (Langfelder–Zhang–Horvath 2008). The main advantage of dynamic cutting compared to the traditional cutting-at-a-specific-level approach is the sesitivity to the shape of the dendogram and to nested groups. Due to our phylogenetic view on BC whereby closer and looser relatedness is assumed to be definitive of groups, we expected nested clusters (that is, groups to be recognized at different levels of cohesion). Therefore, this tool seemed to suit our needs quite well.

**Results and discussion**

Having defined age-sensitive bibliographic coupling (*asBC*) on the basis of the classical approach (*cBC*), we subjected our corpus collected on the history of the species problem to a dual analysis. For the purposes of comparison, we applied both the classical, and the new method to reveal its cognitive structure. In what follows, the results of the two clustering exerciseses are presented and compared. According to our goals, we contrast the respective outputs (1) in a quantitative and (2) in a qualitative manner as well. The qualitative approach, in particular, the thematic characterization of the document clusters yielded by the *asBC* provides, as a demonstration of the capacities of *asBC*, insight into the „historically informed" structure of the species problem.

*Quantitative comparison*

To the effect of a first diagnosis to see whether the results of *cBC* and *asBC* could be expected to show a different picture of the corpus, the degree of similarity between the two groupings were estimated. We used two indicators thereof, (1) the Jaccard index plus (2) the correlation of cophenetic distances within the respective clusterings. The Jaccard index, in this case, could be interpreted as the relative extent of overlap between the two clusterings with a range of values [0,1], and yielded a value of *J* = 0.3, reporting a relatively small portion of document pairs that are judged similarly by both methods. Indicator (2) goes beyond this level of granularity, as it measures the change of relative positions each document has in the cluster tree based on *cBC*, when recalculated via *asBC*. The correlation obtained was *r* = 0.66, indicating that the distances of documents within the cluster tree has moderately changed due to the age-sensitive grouping, that is, groups of documents are more closely or loosely connected on the new account (within in the hierarchical cluster tree). This observation is in accord with our expectations outlined in the previous section. In sum, the two diagnostics suggested that the age-sensitive version of BC generated a refined cognitive structure with different clusters, resulting mainly from the redefinition of document similarity increased or decreased as a function of the age distribution of references.

In mor detail, the classical procedure, *cBC* resulted in a corpus divided into N=4 clusters, while the age-sensitive version, *asBC* yielded N=6 clusters. These numbers, already at this quite general level, suggest that *asBC* did result in a refinement of the clusters from *cBC*. This assumption is further corroborated by the size of these groups (that can be read off from Table 1, see below). While in the original case (cBC), 63% of the sample documents formed a single category, the age-sensitive version produced a more even, less uniform distribution with the first two groups accountig for 36% and 30% of the corpus, respectively. The remaining asBC-clusters were also in a par with the remaining cBC-clusters, that is, no degradation of group size according to the refined method could be observed (indicating small, less „proper" groups, outliers etc.).

**Table 1.** *Comparison of the clusterings obtained by* cBC *vs.* asBC *via a confusion matrix.*

| cBC / asBC | | 1 | 2 | 3 | 4 | 5 | 6 | Sum | (×100) % |
|---|---|---|---|---|---|---|---|---|---|
| | 1 | 89 | 107 | 0 | 27 | 22 | 0 | 245 | 0.63 |
| | 2 | 0 | 0 | 28 | 0 | 0 | 0 | 28 | 0.07 |
| | 3 | 17 | 0 | 0 | 0 | 4 | 0 | 21 | 0.05 |
| | 4 | 34 | 9 | 5 | 4 | 1 | 39 | 92 | 0.24 |
| Sum | | 140 | 116 | 33 | 31 | 27 | 39 | 386 | 1.00 |
| (×100) % | | 0.36 | 0.30 | 0.09 | 0.08 | 0.07 | 0.10 | 1.00 | |

To put it another way, the new method seemed to split the largest (and, as unifying most documents, supposedly somewhat meaningless or hardly interpretable) cluster into smaller ones, that are expected to be historically more coherent (see the qualitative section below). Indeed, the so-called confusion matrix of the two groupings has the same implication (Table 1.). The confusion matrix is a cross-table of the two clusterings, reporting the joint distribution of sample documents within both sets of clusters (so that the relation of *cBC*- and *asBC*-groups could be examined). The rows of Table 1 correspond to the four clusters drawn via the cBC-method, as the columns to the six new clusters from the asBC-method.

As is apparent in the matrix, most affected by the re-partitioning of the species problem literature is the *cBC*-cluster no. 1, that has been split into mainly two, similar-sized groups, *asBC*-clusters no. 1 and no 2. These are also the dominant groups in the matrix, in terms of size. The classical cluster no. 2 and no. 3 remained mostly unchanged, indicating a strong historical–thematic cohesion. Much less robust is the classical cluster 4, similar to no. 1, as its content has also been re-allocated between, primarily, the first and the last age-sensitive cluster (no. 1 and no. 6), but with less constituent elements than the first cluster, altogether.

*Qualitative characterization of new clusters*

According to our primary interest in applying the *asBC* method to the historical corpus in the focus of this study, we also investigated the content of the resulting document clusters, in relation to the classical ones. To this effect, we followed a strategy based on two pillars:

1) Since mapping the intellectual structure of the topic was modelled via references, for the qualitative characterization of these clusters we also relied on the contribution of references to the formation of clusters.
2) In order to obtain a mapping in an economic way, that reveals both the profile of the new clustering and the difference between the „old" and the „new" profile, we did not aim to describe all groups. Instead, we selected a set of clusters that best represented this two aspects at once.

Point 1) above has been addressed by the following procedure: for each selected cluster $C$ the references of documents belonging to $C$ were collected and ranked, according to their cumulative weight in $C$ (that is, their weight used by the *asBC* method times the number of documents they referred by, within $C$). Note that such a cumulative weight is proportional to the contribution of the particular reference to the formation of $C$. In other words, this ranking shows how important a particular reference in the intellectual background of $C$ is. Based on this ranking, we obtained the first $n$ most important reference in $C$ to draw the profile of the cluster. The threshold $n$ was based on a „knee plot" of ranks: the weight-based ordering of reference sets in each case led to a typical powerlaw-like curve with a relatively few references—with high cumulative weight— playing a major role, and many more contributing to a much lower level in itself. We identified these highly-weighted refs as residing in the first, most rapidly ascending section of the weight-curve that ends with a change of slope, the so-called „knee" that can be seen as a transition to the almost flat section of the curve. As the most important descriptors of $C$, we called this $n$ refrences (above the knee of the curve) as the *core* of $C$. In what follows, beyond its description, the core is presented for each cluster under consideration as a set of references, and supported by the knee plot of the cluster. The knee plots are presented in the Appendix, under Fig 4. Core references are also included in the Appendix for each cluster, in the form of ranked lists, collected in Table 2.

Point 2) of our strategy was achieved by selecting *asBC* clusters no. 1–4 to look after contentwise, together with their two sublcusters. One of these is (1) the fragment of no. 1 that previously belonged to the classical (*cBC*-) cluster 1, referred to as 1/1, and (2) another fragment of no. 1 that previously was part of the classical cluster 4, referred to as 1/4. The explanation of this choice leads back to Table 1. It can be seen that by examining *asBC* clusters no. 1 and 2, we can gain insight to the two dominant clusters (in terms of size) of the new thematic profile. On the other hand, since the vast majority of the first *cBC*-cluster has been reallocated between these two and, in addition, no. 4, we may also observe how the oversized „old" thematic group (no. 1) has been reconceptualized by the age-sensitive method. The two sublcusters 1/1 and 1/4 further refine this picture, as while new clusters no. 2 and no. 4 were born almost exclusively from the classical no. 1, new cluster 1 also inherited from old cluster 4. Finally, new cluster no. 3 is discussed as left rather intact (being almost identical to old cluster no. 2). In sum, by this selection, both novel and unchanged parts of the new profile are sampled (*asBC*-clusters 1–2–4 and 3, respectively), and also the relation of the two clusterings may become visible.

Based on these considerations, the historically informed structure of the species problem can be described with the following profiles:

- Cluster no. 1: the BSC and the debate over the theory framing the species concept

The core of the first cluster contains approx. 50 important references, ranked with their cumulative weights in Table 2[1]. (the knee plot on Fig. 4 suggested a threshold of cumulative document weight, CDM > 150). Highly-ranked references are the position papers on the species concept since the modern synthesis. Most striking, especially from the full list of core references including books and book chapters as well, is the dominance of Ernst Mayr, the champion of the „biological species concept" or the BSC (cf. Mayr, #4) what, basically, launched this debate in the context of the synthesis. Several position papers, upon debating the BSC, ranked high in this list. These papers are also classical proposals of infamous alternative species conceptions (not just definitions), such as the „pluralistic conception" or „species pluralism" (Mishler, #10), the „evolutionary species concept" (Wiley, #13), the „genetic species concept" (Masters, #15). With somewhat lower weights, but two further definitions also exhibit themselves, namely, the „phylogenetic concept" (Nixon, #18), and the „ecological species concept" (van Valen, #30), though the latter having the lowest rank in the list.

Beside the collection of proposals to challenge the BSC as the concept that initiated the discourse, a further line of research also observable in Cluster 1, as heavily interacting with the previous one. Among highly ranked papers we find several approaches regarding the application, or, rather, the problems of application of the biological concept (BSC), mainly in microbiology (Wayne, #3; Dykhuizen, #8; Smith, #11 or, as a case outside microbiology, Knowlton, #20). The association of these topics is well-explained by the fact that the BSC is known as hardly applicable to biological kinds with non-sexual reproduction, such as bacteria and other subjects of microbiology, but also has strange implications to some sexually reproducing kinds as well (e.g. sibling species, Knowlton, #20). What we see in this reference set, then, is best interpreted as a series of responses to the BSC on the part of the practice of systematics.

In sum, Cluster no. 1 can be conceived as quite coherently mirroring what is the bottomline of the XX. centrury history of the problem, the biological conception (BSC) and the immediate discourse it generated, including both the application and the alternatives of this concept. In terms of the history and philosophy of biology, this profile is the debate over the best theory of species within biology, yielding a theoretically sound category.

---

[1] In Table 2 only journal publications are demonstrated, therefore, the actual number of references included in the table is smaller than the size of the whole core.

- Cluster no. 2: A more recent response: cladistics and the PSC

The core of the second most extensive cluster counts about 100 references (by the knee plot, CDM > 150, as above). Thematically, this group of referred papers is rather coherent. By inspecting the list, striking is the dominance of two concepts, „cladistics" and the „phylogenetic species concept": at least one parameter of each document is related to one of these notions. Many highly ranked references came from the journal *Cladistics*, which has been the main platform of a specific school of systematics by the same name. The reference of the highest rank is Nixon's seminal paper, published in *Cladistics* on the phylogenetic species concept (#1)—this very paper occured in Cluster no 1 also, but with a relatively low rank, indicating a different emphasis of the two clusters. Papers from other journals also contain „cladistics" and/or a reference to the phylogenetic species concept in their metadata, among their keywords or within their abstracts, with a very few exception. The unity of the profile is also confirmed by the ISI Subject Categories assigned to the papers included: almost each assignment contains „Evolutionary Biology", and, in the majority of the cases, quite exclusively.

Due to this relatively clear profile, Cluster no 2. can be interpreted as the „cladistic response" to the species problem (or, to the BSC). Cladistics is a more recent development in systematics, a school with very specific implications on the definition of the species category, concerning how the phylogenetic tree should be partitioned into species. It is, therefore, closely related to the so-called „phylogenetic species concept" (PSC). The representation of this school is also expressed by the high rank and recurrence of a set of authors, known as the champions of either the phylogenetic or the cladistic conceptualization, e.g. Donoghue, DeQueiroz, Cracraft, Mishler etc. In sum, the cluster is a body of literature on this school of systematics entering the species broblem, and producing a significant part of its history.

- Cluster no. 3: the species problem in ecology—a thematic outlier

The core of the *asBC*-cluster no. 3 is a relatively small one, enumerating 15 important references altogether (CDM > 150). Characteristic of its thematic composition are two features of the document set: (1) the references of the two (or three) highest rank are far above the others in terms of weight, and are concerned with the „keystone species concept" (in ecology), and (2) the Subject Category to which these pubs have been assigned by WoS is mainly *Ecology* (and rarely is *Evolutionary Biology*, as opposed to the previous clusters).

This rather compact thematic group is an interesting example of what can be called a „thematic outlier", a strain of research that doesn't belong to the (history of the) very problem under study. Being a „self-contained" group is also reflected in the robustness of the cluster: as noted above, both methods, *cBC* and *asBC* classified these refences nearly the same way, as cluster 3 was originated from classical cluster 2 almost without any change (cf. Table 1).

The reason for this sub-topic entering our sample can be said mainly terminological: both discourses are called „species problem" in their own (otherwise, related) contexts. However, while our interest lies in the discourse on the appropriate species concept for biology, the more particular discourse indicated here belongs to the field of ecology and addresses the role of species as actors setting up ecosystems. Therefore, while in the former case the „species problem" stands for the problem of the species concept, in the latter it denotes the problem of finding species in ecosystems (e.g. foodwebs) whose presence are crucial for its functioning (keystone species). Consequently, in this case, the method (actually, both methods) of bibliographic coupling can be credited for „filtering out" a direction that doesn't belong to the scope of the study.

- Cluster no. 4: An ontology of species taxa for the theory of species

The new cluster no. 4 is also based on a relatively small core, containing about 20 references. The threshold level, CDM > 100, drawn from the knee plot is below the level encountered for the previous clusters, indicating that it is a somewhat less coherent, or more diverse intellectual basis compared to those of the other three groups. A quite interesting multi- (or, as we shall see, rather inter-) disciplinarity can also be observed as to the thematic structure: The pub of the highest rank (ref1) refers to the solution of cladistics to the species problem, yet is has been published in the journal *Biology and Philosophy*, wich fact is also reflected in its Subject Category, *History & Philosophy of Science*. This very Subject Category dominates a significant part of the core, together with *Zoology*. What this mixture of „cultures" conveys is a very authentic feature of the species problem, well represented in this separate cluster.

The feature in question is a clear tendency within the XX. century scientific debate on species to rely on and properly icorporate arguments from the philosophy of science (namely, of biology). Just as Darwin revolutionalized systematics by altering the way we look at individual species (species taxa), so did, in the modern history of the problem, two authors, Micheal Ghiselin (a biologist) and David Hull (a philosopher of science), the champions of the „individuality thesis" (species as individuals, SAI). Addressing the ontology of species (taxa), they argued that species are best viewed, instead of being „classes of organisms", as individuals (particular, historical, evolvable etc. entities). Interestingly, in the technical sense, this view supported some definitions of species, while discrediting others. Among those that could directly rely on SAI was the cladistic species concept and its relatives. As a result of the interaction between biophilosophy and systematics, the SAI and other ontological arguments became integral part of the scientific discourse on species.

This quick historical highligt makes cluster no. 4 a well-interpretable collection. Authors of this cluster are, indeed Ghiselin, Hull and other theoreticians and biophilosophers (Kitcher, Kluge), on one hand, and proponents of the cladistic and phylogenetic concept, on the other (Ridley, DeQuerioz, Mishler, Cracraft etc.). Beyond the synbiontic relation of these two cultures, the presence of the practice of systematics is also present with a high rank (#2). This indicates that theorizing on the status of species propagated into the very circles of practitioners of systematics as well. In sum, cluster 4 can be conceived as a trace of the debate on the ontology of species taxa, being infiltrated into biological theorizing about the species concept (category).

- Cluster no. 1/1 and 1/4: Aquiring historical coherence

The remaining two groups we took under closer inspection were both a fragment of no. 1 described above. The main reason for looking into the internal structure of the first cluster was to sharpen the characterization of how the age-sensitive restructuring of the corpus affected the original thematic groups.

Cluster 1/1 is the fraction of our new cluster no. 1 (The BSC-related theme), that came from the original cluster 1. Recall, that the striking change from the re-clustering procedure was the division of old cluster 1 into new ones, exposed so far as the new cluster no. 1 (condisering the majority of its content) and 2. However, it is somewhat more sound to speak of new cluster 1/1 and 2 as the resulting groups. Now, by turning to the content of 1/1, we encounter an even more concentrated profile, than that of the whole class: in this fragment, the the position papers proposing and discussing the BSC and its major alternatives exhibit themselves, that is, theorizing of the main figures of biosystematics about the species concept (category). Even more telling, with respect to the capacity of the age-sensitive method, if we compare the age distribution of references in cluster 1/1 and 2, respectively, that is, between the two descendant of the same old cluster. According to Fig 3, the *asBC* procedure sorted the content of the old cluster into a „more classical", and a „more recent" discussion. For cluster 1/1, references are distributed almost equally before and after the '90s, with a peak in the late 80's, while for cluster 2 the majority of references originate from the '90s, their peak is in the early '90s, and show a more „continuous" or coherent discourse. In other words, the procedure identified the BSC-based dispute (cluster 1/1) as a more classical context, within which the new cluster no. 2, that is, the cladistic/phylogenetic discourse emerged as a more recent movement. Note, that these two, historically distinguishable movements were inseparably linked together by the *cBC* method, in one, thematically coherent but giant cluster. In this sense, the *asBC* method did produce a historically informed thematic structure, differentiating between „ancient" and „new" features of a thematic group.

Considering the contribution of 1/4, the fraction of the BSC-theme that came from the classical cluster no. 4, the picture gets even more interesting. In this small fragment (the core contains only 12 pubs) papers (references) from the very practice of biosystematics are added to the theoretical debate in 1/1, belonging, in particular, to the field of microbiology. This phenomenon recalls our previous observation that new cluster no. 1 covers both (1) the theoretical debate initiated by the biological species concept (BSC) and (2) its extension from, mainly, microbiology, whereby the application of BSC has always been problematic. At this point, we can see that not only does this cluster unify these references, but also „collects" them by „cutting out" the theoretical and the applied part of the BSC-debate from old clusters 1 and 4.

In sum, results suggest that the proposed method of *asBC* has been capable of better identifying strains of research or schools in the modern history of the species problem.

On one hand, the *asBC* eliminated a more recent school within the theoretical discourse, namely, the phylogenetic approach and cladistics emerging from the pool of species concepts. On the other hand, it unified references that show the real or causal, that is, historical unfolding of ideas, instead of reflecting mere topical similarities. This latter feature is shown in connecting the theory and application of the BSC, while, in the original cluster structure these pubs were sorted into the big „theoretical cluster" (old cluster 1), and the „cluster of applications", mainly, topics in microbiology (old cluster 4), respectively.

**Fig. 3.** *Age distribution of references within the core of clusters 1/1 and 2, respectively.*

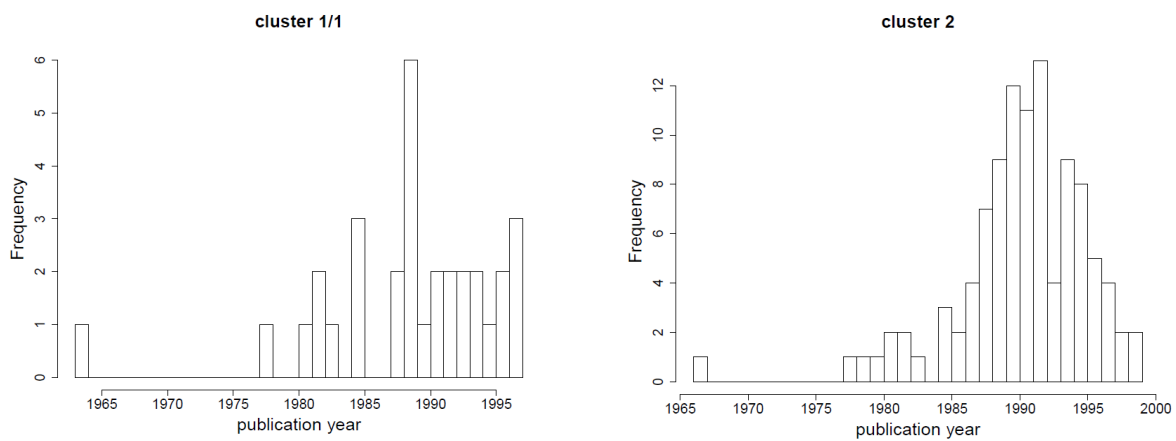

**Conclusion**

In this paper, we proposed a method of bibliographic coupling (BC) designed primarily for the purposes of the history of science. As an alternative of classical BC, age-sensitive bibliographic coupling, or *asBC*, was supposed to work in a similar manner as evolutionary systematics does in biology. By incorporating the age (in the bibliometrics case: publication year) of common ancestors (references) into the assessment of document relatedness, it was supposed to support a classification of source documents that reflects the history of the subject. Not only were the resulting clusters expected to distinguish the various research directions emerged within the area under study, but also to mirror the historical relations between these directions.

Having defined the age-sensitive method, we applied it in a pilot bibliometric study of an important decade of the Species Problem, a centuries-old but still active discourse in biology addressing the concept of biological species. Quantitative results showed that the new method was able to refine the thematic structure of the corpus, collected on the Species Problem, that was obtained by the classical method of bibliographic coupling. The comparison of the two clusterings (the classical and the altered one) made clear that the giant thematic cluster, resulted from classical BC, was split up by the age-sensitive method.

Truly promising observations were gained on closer inspection of the document clusters resulted from *asBC*, that is, from the qualitative assessment and comparison of the new structure vs. the classical one. Via the age-sensitive method the extensive theoretical debate on species, classified in the giant cluster mentioned above by the original method of BC, could be differentiated into clusters representing the initial context of the discourse, and later developments, like coherent schools of systematics responding to the initial context. On the other hand, documents differing topicwise, but belonging to the same research tradition were bound together by the new method, while this relation was overlooked in the classical case.

In sum, the method of *asBC* seemed to be a utility that is worth experimentig with. As a procedure for detecting either research dynamics or patterns in the history of science, age-sensitive bibliometric coupling could be a useful tool for bibliometric investigations aiding the historian of complex scientific discourses. Subsequent research is intended to work on the refinement of this measure, as well as on further clarification, via bibliometric means, of the historical structure of the Species Problem. Iteratively contrasting these two would result in an efficient empirical methodology for mapping complicated historical phenomena in science.

**Acknowledgement**

This paper was supported by the János Bolyai Research Scholarship of the Hungarian Academy of Sciences.

# Appendix

**Fig 4.** *The „knee plots" of clusters 1–4, respectively, supporting the extraction of core references for each. Cumulative weights are plotted against the indices of ranked references. Only the section of the whole curve is graphed where its „knee" is observable.*

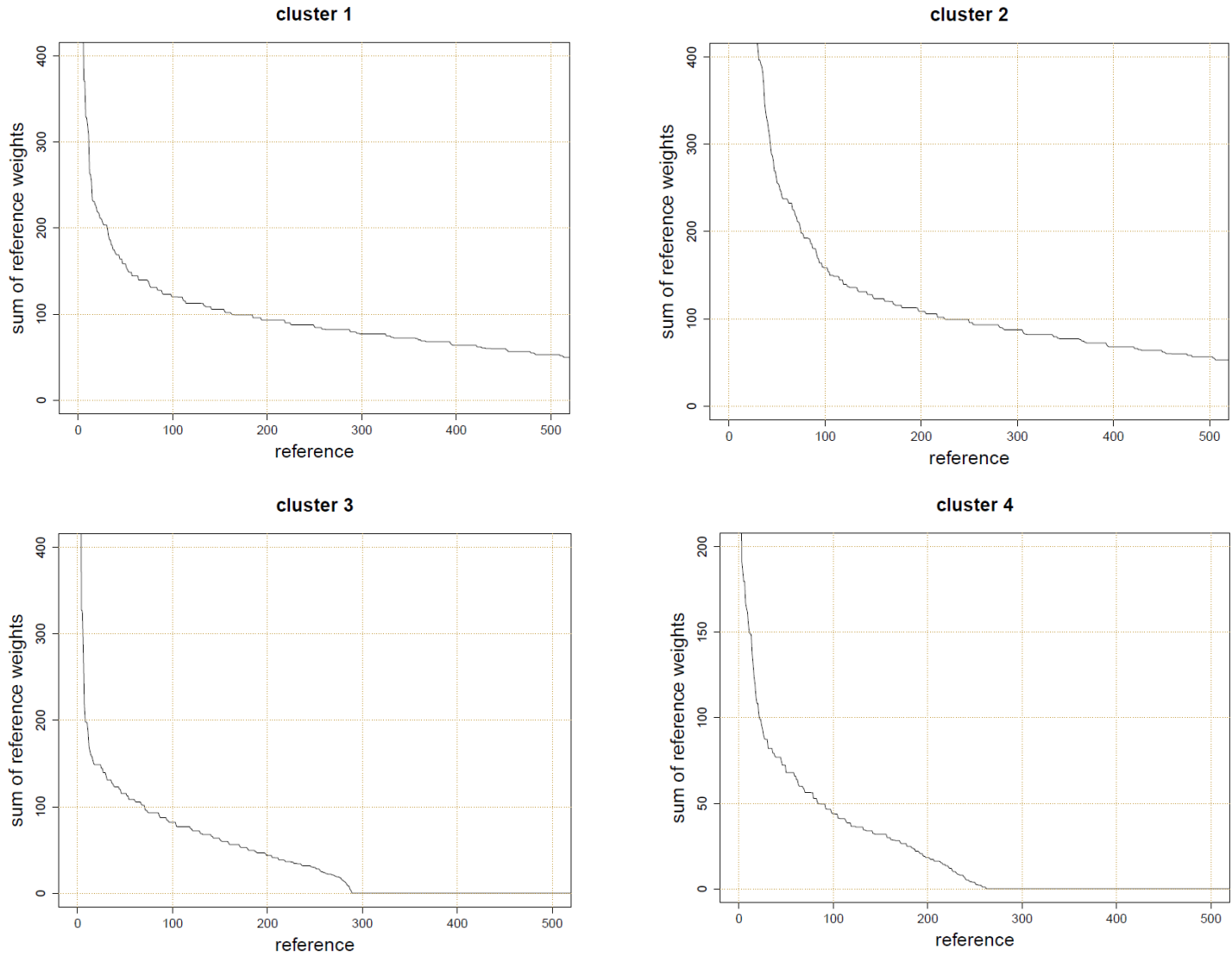

**Table 2.** *The lists of core references within asBC-clusters 1–4, respectively. In this excerpt, only journal publications are listed. Items are ranked according to their cumulative weight, referred by „Sum of weights" (age-related weight of reference* R *× number of occurences of reference* R *within the cluster).*

## CLUSTER 1

| # | Reference (WoS format) | Sum of weights | Title | WoS Category (Subject Category) |
|---|---|---|---|---|
| 1 | MALLET J, 1995, TRENDS ECOL EVOL, V10, P294 | 644,56 | A SPECIES DEFINITION FOR THE MODERN SYNTHESIS | Ecology; Evolutionary Biology; Genetics & Heredity |
| 2 | COYNE JA, 1988, SYST ZOOL, V37, P190 | 446,37 | DO WE NEED A NEW SPECIES CONCEPT | Zoology |
| 3 | WAYNE LG, 1987, INT J SYST BACTERIOL, V37, P463 | 329,43 | REPORT OF THE AD-HOC-COMMITTEE ON RECONCILIATION OF APPROACHES TO BACTERIAL SYSTEMATICS | Microbiology |
| 4 | MAYR E, 1992, AM J BOT, V79, P222 | 328,07 | A LOCAL FLORA AND THE BIOLOGICAL SPECIES CONCEPT | Plant Sciences |
| 5 | Mann DG, 1996, HYDROBIOLOGIA, V336, P19 | 264,15 | BIODIVERSITY, BIOGEOGRAPHY AND CONSERVATION OF DIATOMS | Marine & Freshwater Biology |
| 6 | VALBONESI A, 1988, J PROTOZOOL, V35, P38 | 255,07 | AN INTEGRATED STUDY OF THE SPECIES PROBLEM IN THE EUPLOTES-CRASSUS-MINUTA-VANNUS GROUP | Zoology |
| 7 | COLEMAN AW, 1994, J PHYCOL, V30, P80 | 232,68 | MOLECULAR DELINEATION OF SPECIES AND SYNGENS IN VOLVOCACEAN GREEN-ALGAE (CHLOROPHYTA) | Plant Sciences; Marine & Freshwater Biology |
| 8 | DYKHUIZEN DE, 1991, J BACTERIOL, V173, P7257 | 231,01 | RECOMBINATION IN ESCHERICHIA-COLI AND THE DEFINITION OF BIOLOGICAL SPECIES | Microbiology |
| 9 | SMITH JM, 1991, NATURE, V349, P29 | 231,01 | LOCALIZED SEX IN BACTERIA | Multidisciplinary Sciences |
| 10 | MISHLER BD, 1982, SYST ZOOL, V31, P491 | 219,46 | SPECIES CONCEPTS - A CASE FOR PLURALISM | Zoology |
| 11 | SMITH JM, 1993, P NATL ACAD SCI USA, V90, P4384 | 218,42 | HOW CLONAL ARE BACTERIA | Multidisciplinary Sciences |
| 12 | GIANNI A, 1990, EUR J PROTISTOL, V26, P142 | 216,91 | AUTOECOLOGICAL AND MOLECULAR APPROACH TO THE SPECIES PROBLEM IN THE EUPLOTES-VANNUS-CRASSUS-MINUTA GROUP (CILIOPHORA, HYPOTRICHIDA) | Microbiology |
| 13 | WILEY EO, 1978, SYST ZOOL, V27, P17 | 206,02 | EVOLUTIONARY SPECIES CONCEPT RECONSIDERED | Zoology |
| 14 | MANN DG, 1989, PLANT SYST EVOL, V164, P215 | 203,69 | THE SPECIES CONCEPT IN DIATOMS - EVIDENCE FOR MORPHOLOGICALLY DISTINCT, SYMPATRIC GAMODEMES IN 4 EPIPELIC SPECIES | Plant Sciences; Evolutionary Biology |
| 15 | MASTERS JC, 1989, SYST ZOOL, V38, P270 | 203,69 | WHY WE NEED A NEW GENETIC SPECIES CONCEPT | Zoology |
| 16 | SCHLEGEL M, 1988, EUR J | 191,3 | TAXONOMY AND PHYLOGENETIC RELATIONSHIP OF 8 SPECIES OF THE GENUS | Microbiology |

| #  | Reference (WoS format) | Sum of weights | Title | WoS Category (Subject Category) |
|----|------------------------|----------------|-------|----------------------------------|
|    | PROTISTOL, V24, P22 |  | EUPLOTES (HYPOTRICHIDA, CILIOPHORA) AS REVEALED BY ENZYME ELECTROPHORESIS |  |
| 17 | CAPRETTE CL, 1994, J EUKARYOT MICROBIOL, V41, P316 | 186,15 | QUANTITATIVE-ANALYSES OF INTERBREEDING IN POPULATIONS OF VANNUS-MORPHOTYPE EUPLOTES, WITH SPECIAL ATTENTION TO THE NOMINAL SPECIES E-VANNUS AND EUPLOTES-CRASSUS | Microbiology |
| 18 | NIXON KC, 1990, CLADISTICS, V6, P211 | 180,76 | AN AMPLIFICATION OF THE PHYLOGENETIC SPECIES CONCEPT | Evolutionary Biology |
| 19 | WOESE CR, 1987, MICROBIOL REV, V51, P221 | 179,69 | BACTERIAL EVOLUTION | Microbiology |
| 20 | KNOWLTON N, 1993, ANNU REV ECOL SYST, V24, P189 | 174,73 | SIBLING SPECIES IN THE SEA | Ecology; Evolutionary Biology |
| 21 | SONNEBORN TM, 1975, T AM MICROSC SOC, V94, P155 | 171,78 | PARAMECIUM-AURELIA COMPLEX OF 14 SIBLING SPECIES | Microscopy |
| 22 | FOX GE, 1992, INT J SYST BACTERIOL, V42, P166 | 164,03 | HOW CLOSE IS CLOSE - 16S RIBOSOMAL-RNA SEQUENCE IDENTITY MAY NOT BE SUFFICIENT TO GUARANTEE SPECIES IDENTITY | Microbiology |
| 23 | GRANT PR, 1992, SCIENCE, V256, P193 | 164,03 | HYBRIDIZATION OF BIRD SPECIES | Multidisciplinary Sciences |
| 24 | VALBONESI A, 1992, J PROTOZOOL, V39, P45 | 164,03 | THE SPECIES PROBLEM IN A CILIATE WITH A HIGH MULTIPLE MATING TYPE SYSTEM, EUPLOTES-CRASSUS | Zoology |
| 25 | BARTON NH, 1985, ANNU REV ECOL SYST, V16, P113 | 158,6 | ANALYSIS OF HYBRID ZONES | Ecology; Evolutionary Biology |
| 26 | FELSENSTEIN J, 1985, EVOLUTION, V39, P783 | 158,6 | CONFIDENCE-LIMITS ON PHYLOGENIES - AN APPROACH USING THE BOOTSTRAP | Ecology; Evolutionary Biology; Genetics & Heredity |
| 27 | Berlocher SH, 1996, HEREDITY, V77, P83 | 158,49 | POPULATION STRUCTURE OF RHAGOLETIS POMONELLA, THE APPLE MAGGOT FLY | Ecology; Evolutionary Biology; Genetics & Heredity |
| 28 | Finlay BJ, 1996, Q REV BIOL, V71, P221 | 158,49 | BIODIVERSITY AT THE MICROBIAL LEVEL: THE NUMBER OF FREE-LIVING CILIATES IN THE BIOSPHERE | Biology |
| 29 | MEDLIN LK, 1991, J PHYCOL, V27, P514 | 154,01 | MORPHOLOGICAL AND GENETIC-VARIATION WITHIN THE DIATOM SKELETONEMA-COSTATUM (BACILLARIOPHYTA) - EVIDENCE FOR A NEW SPECIES, SKELETONEMA-PSEUDOCOSTATUM | Plant Sciences; Marine & Freshwater Biology |
| 30 | VANVALEN L, 1976, TAXON, V25, P233 | 152,06 | ECOLOGICAL SPECIES, MULTISPECIES, AND OAKS | Plant Sciences; Evolutionary Biology |

## CLUSTER 2

| #  | Reference (WoS format) | Sum of weights | Title | WoS Category (Subject Category) |
|----|------------------------|----------------|-------|----------------------------------|

| # | Reference (WoS format) | Sum of weights | Title | WoS Category (Subject Category) |
|---|---|---|---|---|
| 1 | NIXON KC, 1990, CLADISTICS, V6, P211 | 2205,23 | AN AMPLIFICATION OF THE PHYLOGENETIC SPECIES CONCEPT | Evolutionary Biology |
| 2 | DAVIS JI, 1992, SYST BIOL, V41, P421 | 1353,28 | POPULATIONS, GENETIC-VARIATION, AND THE DELIMITATION OF PHYLOGENETIC SPECIES | Evolutionary Biology |
| 3 | DONOGHUE MJ, 1985, BRYOLOGIST, V88, P172 | 1321,69 | A CRITIQUE OF THE BIOLOGICAL SPECIES CONCEPT AND RECOMMENDATIONS FOR A PHYLOGENETIC ALTERNATIVE | Plant Sciences |
| 4 | DEQUEIROZ K, 1988, CLADISTICS, V4, P317 | 1275,35 | PHYLOGENETIC SYSTEMATICS AND THE SPECIES PROBLEM | Evolutionary Biology |
| 5 | BAUM DA, 1995, SYST BOT, V20, P560 | 644,56 | CHOOSING AMONG ALTERNATIVE PHYLOGENETIC SPECIES CONCEPTS | Plant Sciences; Evolutionary Biology |
| 6 | DEQUEIROZ K, 1990, CLADISTICS, V6, P61 | 614,57 | PHYLOGENETIC SYSTEMATICS OR NELSONS VERSION OF CLADISTICS | Evolutionary Biology |
| 7 | WHEELER QD, 1990, CLADISTICS, V6, P77 | 614,57 | ANOTHER WAY OF LOOKING AT THE SPECIES PROBLEM - A REPLY TO DEQUEIROZ AND DONOGHUE | Evolutionary Biology |
| 8 | DEQUEIROZ K, 1990, CLADISTICS, V6, P83 | 578,42 | PHYLOGENETIC SYSTEMATICS AND SPECIES REVISITED | Evolutionary Biology |
| 9 | MALLET J, 1995, TRENDS ECOL EVOL, V10, P294 | 545,4 | A SPECIES DEFINITION FOR THE MODERN SYNTHESIS | Ecology; Evolutionary Biology; Genetics & Heredity |
| 10 | DAVIS JI, 1991, SYST BOT, V16, P431 | 539,02 | ISOZYME VARIATION AND SPECIES DELIMITATION IN THE PUCCINELLIA-NUTTALLIANA COMPLEX (POACEAE) - AN APPLICATION OF THE PHYLOGENETIC SPECIES CONCEPT | Plant Sciences; Evolutionary Biology |
| 11 | CRACRAFT J, 1992, CLADISTICS, V8, P1 | 533,11 | THE SPECIES OF THE BIRDS-OF-PARADISE (PARADISAEIDAE) - APPLYING THE PHYLOGENETIC SPECIES CONCEPT TO A COMPLEX PATTERN OF DIVERSIFICATION | Evolutionary Biology |
| 12 | OHARA RJ, 1993, SYST BIOL, V42, P231 | 524,2 | SYSTEMATIC GENERALIZATION, HISTORICAL FATE, AND THE SPECIES PROBLEM | Evolutionary Biology |
| 13 | NELSON G, 1989, CLADISTICS, V5, P275 | 509,23 | CLADISTICS AND EVOLUTIONARY MODELS | Evolutionary Biology |
| 14 | BAUM D, 1992, TRENDS ECOL EVOL, V7, P1 | 492,1 | PHYLOGENETIC SPECIES CONCEPTS | Ecology; Evolutionary Biology; Genetics & Heredity |
| 15 | DOYLE JJ, 1992, SYST BOT, V17, P144 | 492,1 | GENE TREES AND SPECIES TREES - MOLECULAR SYSTEMATICS AS ONE-CHARACTER TAXONOMY | Plant Sciences; Evolutionary Biology |
| 16 | MCKITRICK MC, 1988, CONDOR, V90, P1 | 478,26 | SPECIES CONCEPTS IN ORNITHOLOGY | Ornithology |
| 17 | FROST DR, 1990, HERPETOLOGICA, V46, P87 | 469,97 | SPECIES IN CONCEPT AND PRACTICE - HERPETOLOGICAL APPLICATIONS | Zoology |
| 18 | VRANA P, 1992, CLADISTICS, V8, P67 | 451,09 | INDIVIDUAL ORGANISMS AS TERMINAL ENTITIES - LAYING THE SPECIES PROBLEM TO REST | Evolutionary Biology |
| 19 | MISHLER BD, 1982, SYST ZOOL, V31, P491 | 438,91 | SPECIES CONCEPTS - A CASE FOR PLURALISM | Zoology |
| 20 | DEQUEIROZ K, 1994, TRENDS | 418,83 | TOWARD A PHYLOGENETIC SYSTEM OF BIOLOGICAL NOMENCLATURE | Ecology; Evolutionary Biology; |

| # | Reference (WoS format) | Sum of weights | Title | WoS Category (Subject Category) |
|---|---|---|---|---|
|  | ECOL EVOL, V9, P27 |  |  | Genetics & Heredity |
| 21 | DEQUEIROZ K, 1992, ANNU REV ECOL SYST, V23, P449 | 410,09 | PHYLOGENETIC TAXONOMY | Ecology; Evolutionary Biology |
| 22 | DOYLE JJ, 1995, SYST BOT, V20, P574 | 396,65 | THE IRRELEVANCE OF ALLELE TREE TOPOLOGIES FOR SPECIES DELIMITATION, AND A NONTOPOLOGICAL ALTERNATIVE | Plant Sciences; Evolutionary Biology |
| 23 | DEQUEIROZ K, 1988, PHILOS SCI, V55, P238 | 382,6 | SYSTEMATICS AND THE DARWINIAN REVOLUTION | History & Philosophy Of Science |
| 24 | MORITZ C, 1994, TRENDS ECOL EVOL, V9, P373 | 325,76 | DEFINING EVOLUTIONARILY-SIGNIFICANT-UNITS FOR CONSERVATION | Ecology; Evolutionary Biology; Genetics & Heredity |
| 25 | AVISE JC, 1987, ANNU REV ECOL SYST, V18, P489 | 299,48 | INTRASPECIFIC PHYLOGEOGRAPHY - THE MITOCHONDRIAL-DNA BRIDGE BETWEEN POPULATION-GENETICS AND SYSTEMATICS | Ecology; Evolutionary Biology |
| 26 | DEQUEIROZ K, 1990, SYST ZOOL, V39, P307 | 289,21 | PHYLOGENY AS A CENTRAL PRINCIPLE IN TAXONOMY - PHYLOGENETIC DEFINITIONS OF TAXON NAMES | Zoology |
| 27 | Avise JC, 1997, P NATL ACAD SCI USA, V94, P7748, DOI 10.1073/pnas.94.15.7748 | 281,47 | PHYLOGENETICS AND THE ORIGIN OF SPECIES | Multidisciplinary Sciences |
| 28 | VANEWRIGHT RI, 1991, BIOL CONSERV, V55, P235 | 269,51 | WHAT TO PROTECT - SYSTEMATICS AND THE AGONY OF CHOICE | Biodiversity Conservation; Ecology; Environmental Sciences |
| 29 | VILGALYS R, 1991, MYCOLOGIA, V83, P758 | 269,51 | SPECIATION AND SPECIES CONCEPTS IN THE COLLYBIA-DRYOPHILA COMPLEX | Mycology |
| 30 | Taylor JW, 1999, CLIN MICROBIOL REV, V12, P126 | 255,74 | THE EVOLUTIONARY BIOLOGY AND POPULATION GENETICS UNDERLYING FUNGAL STRAIN TYPING | Microbiology |
| 31 | PAMILO P, 1988, MOL BIOL EVOL, V5, P568 | 255,07 | RELATIONSHIPS BETWEEN GENE TREES AND SPECIES TREES | Biochemistry & Molecular Biology; Evolutionary Biology; Genetics & Heredity |
| 32 | CHASE TE, 1990, MYCOLOGIA, V82, P67 | 253,06 | GENETIC-BASIS OF BIOLOGICAL SPECIES IN HETEROBASIDION-ANNOSUM - MENDELIAN DETERMINANTS | Mycology |
| 33 | GRAYBEAL A, 1995, SYST BIOL, V44, P237 | 247,91 | NAMING SPECIES | Evolutionary Biology |
| 34 | DEQUEIROZ K, 1992, BIOL PHILOS, V7, P295 | 246,05 | PHYLOGENETIC DEFINITIONS AND TAXONOMIC PHILOSOPHY | History & Philosophy Of Science |
| 35 | Geiser DM, 1998, P NATL ACAD SCI USA, V95, P388, DOI 10.1073/pnas.95.1.388 | 239,97 | CRYPTIC SPECIATION AND RECOMBINATION IN THE AFLATOXIN-PRODUCING FUNGUS ASPERGILLUS FLAVUS | Multidisciplinary Sciences |
| 36 | AVISE JC, 1989, EVOLUTION, V43, P1192 | 237,64 | GENE TREES AND ORGANISMAL HISTORIES - A PHYLOGENETIC APPROACH TO POPULATION BIOLOGY | Ecology; Evolutionary Biology; Genetics & Heredity |
| 37 | KLUGE AG, 1989, CLADISTICS, V5, P291 | 237,64 | METACLADISTICS | Evolutionary Biology |
| 38 | RIDLEY M, 1989, BIOL PHILOS, V4, P1 | 237,64 | THE CLADISTIC SOLUTION TO THE SPECIES PROBLEM | History & Philosophy Of Science |

| # | Reference (WoS format) | Sum of weights | Title | WoS Category (Subject Category) |
|---|---|---|---|---|
| 39 | FROST DR, 1994, CLADISTICS, V10, P259 | 232,68 | A CONSIDERATION OF EPISTEMOLOGY IN SYSTEMATIC BIOLOGY, WITH SPECIAL REFERENCE TO SPECIES | Evolutionary Biology |
| 40 | OHARA RJ, 1994, AM ZOOL, V34, P12 | 232,68 | EVOLUTIONARY HISTORY AND THE SPECIES PROBLEM | Zoology |
| 41 | PATTON JL, 1994, SYST BIOL, V43, P11 | 232,68 | PARAPHYLY, POLYPHYLY, AND THE NATURE OF SPECIES BOUNDARIES IN POCKET GOPHERS (GENUS-THOMOMYS) | Evolutionary Biology |
| 42 | VOGLER AP, 1994, CONSERV BIOL, V8, P354 | 232,68 | DIAGNOSING UNITS OF CONSERVATION MANAGEMENT | Biodiversity Conservation; Ecology; Environmental Sciences |
| 43 | Koufopanou V, 1997, P NATL ACAD SCI USA, V94, P5478, DOI 10.1073/pnas.94.10.5478 | 225,18 | CONCORDANCE OF GENE GENEALOGIES REVEALS REPRODUCTIVE ISOLATION IN THE PATHOGENIC FUNGUS COCCIDIOIDES IMMITIS | Multidisciplinary Sciences |
| 44 | BOIDIN J, 1986, MYCOTAXON, V26, P319 | 225,07 | INTERCOMPATIBILITY AND THE SPECIES CONCEPT IN THE SAPROBIC BASIDIOMYCOTINA | Mycology |
| 45 | WILEY EO, 1978, SYST ZOOL, V27, P17 | 223,19 | EVOLUTIONARY SPECIES CONCEPT RECONSIDERED | Zoology |
| 46 | KORNET DJ, 1993, J THEOR BIOL, V164, P407 | 218,42 | PERMANENT SPLITS AS SPECIATION EVENTS - A FORMAL RECONSTRUCTION OF THE INTERNODAL SPECIES CONCEPT | Biology; Mathematical & Computational Biology |
| 47 | VILGALYS R, 1990, J BACTERIOL, V172, P4238 | 216,91 | RAPID GENETIC IDENTIFICATION AND MAPPING OF ENZYMATICALLY AMPLIFIED RIBOSOMAL DNA FROM SEVERAL CRYPTOCOCCUS SPECIES | Microbiology |
| 48 | FELSENSTEIN J, 1985, EVOLUTION, V39, P783 | 211,47 | CONFIDENCE-LIMITS ON PHYLOGENIES - AN APPROACH USING THE BOOTSTRAP | Ecology; Evolutionary Biology; Genetics & Heredity |
| 49 | Huelsenbeck JP, 1996, TRENDS ECOL EVOL, V11, P152 | 211,32 | COMBINING DATA IN PHYLOGENETIC ANALYSIS | Ecology; Evolutionary Biology; Genetics & Heredity |
| 50 | MAYR E, 1992, AM J BOT, V79, P222 | 205,04 | A LOCAL FLORA AND THE BIOLOGICAL SPECIES CONCEPT | Plant Sciences |
| 51 | LUCKOW M, 1995, SYST BOT, V20, P589 | 198,33 | SPECIES CONCEPTS - ASSUMPTIONS, METHODS, AND APPLICATIONS | Plant Sciences; Evolutionary Biology |
| 52 | FARRIS JS, 1991, CLADISTICS, V7, P297 | 192,51 | HENNIG DEFINED PARAPHYLY | Evolutionary Biology |
| 53 | HARRISON RG, 1991, ANNU REV ECOL SYST, V22, P281 | 192,51 | MOLECULAR-CHANGES AT SPECIATION | Ecology; Evolutionary Biology |
| 54 | Kasuga T, 1999, J CLIN MICROBIOL, V37, P653 | 191,81 | PHYLOGENETIC RELATIONSHIPS OF VARIETIES AND GEOGRAPHICAL GROUPS OF THE HUMAN PATHOGENIC FUNGUS HISTOPLASMA CAPSULATUM DARLING | Microbiology |
| 55 | CODDINGTON JA, 1988, CLADISTICS, V4, P3 | 191,3 | CLADISTIC TESTS OF ADAPTATIONAL HYPOTHESES | Evolutionary Biology |
| 56 | FARRIS JS, 1994, CLADISTICS, V10, P315 | 186,15 | TESTING SIGNIFICANCE OF INCONGRUENCE | Evolutionary Biology |
| 57 | MORITZ C, 1994, MOL ECOL, V3, P401 | 186,15 | APPLICATIONS OF MITOCHONDRIAL-DNA ANALYSIS IN CONSERVATION - A CRITICAL-REVIEW | Biochemistry & Molecular Biology; Ecology; Evolutionary |

| # | Reference (WoS format) | Sum of weights | Title | WoS Category (Subject Category) |
|---|---|---|---|---|
| | | | | Biology |
| 58 | CHASE TE, 1990, MYCOLOGIA, V82, P73 | 180,76 | 5 GENES DETERMINING INTERSTERILITY IN HETEROBASIDION-ANNOSUM | Mycology |
| 59 | KLUGE AG, 1990, BIOL PHILOS, V5, P417 | 180,76 | SPECIES AS HISTORICAL INDIVIDUALS | History & Philosophy Of Science |
| 60 | O'Donnell K, 1998, MYCOLOGIA, V90, P465 | 179,97 | MOLECULAR SYSTEMATICS AND PHYLOGEOGRAPHY OF THE GIBBERELLA FUJIKUROI SPECIES COMPLEX | Mycology |
| 61 | BAKER CS, 1993, P NATL ACAD SCI USA, V90, P8239 | 174,73 | ABUNDANT MITOCHONDRIAL-DNA VARIATION AND WORLDWIDE POPULATION-STRUCTURE IN HUMPBACK WHALES | Multidisciplinary Sciences |
| 62 | CHAPPILL JA, 1989, CLADISTICS, V5, P217 | 169,74 | QUANTITATIVE CHARACTERS IN PHYLOGENETIC ANALYSIS | Evolutionary Biology |
| 63 | Burt A, 1997, MOL ECOL, V6, P781, DOI 10.1046/j.1365-294X.1997.00245.x | 168,88 | MOLECULAR MARKERS REVEAL DIFFERENTIATION AMONG ISOLATES OF COCCIDIOIDES IMMITIS FROM CALIFORNIA, ARIZONA AND TEXAS | Biochemistry & Molecular Biology; Ecology; Evolutionary Biology |
| 64 | HILLIS DM, 1992, J HERED, V83, P189 | 164,03 | SIGNAL, NOISE, AND RELIABILITY IN MOLECULAR PHYLOGENETIC ANALYSES | Genetics & Heredity |
| 65 | ROJAS M, 1992, CONSERV BIOL, V6, P170 | 164,03 | THE SPECIES PROBLEM AND CONSERVATION - WHAT ARE WE PROTECTING | Biodiversity Conservation; Ecology; Environmental Sciences |
| 66 | COYNE JA, 1988, SYST ZOOL, V37, P190 | 159,42 | DO WE NEED A NEW SPECIES CONCEPT | Zoology |
| 67 | OHARA RJ, 1988, SYST ZOOL, V37, P142 | 159,42 | HOMAGE TO CLIO, OR, TOWARD AN HISTORICAL PHILOSOPHY FOR EVOLUTIONARY BIOLOGY | Zoology |
| 68 | Burt A, 1996, P NATL ACAD SCI USA, V93, P770 | 158,49 | MOLECULAR MARKERS REVEAL CRYPTIC SEX IN THE HUMAN PATHOGEN COCCIDIOIDES IMMITIS | Multidisciplinary Sciences |
| 69 | Legge JT, 1996, CONSERV BIOL, V10, P85 | 158,49 | GENETIC CRITERIA FOR ESTABLISHING EVOLUTIONARILY SIGNIFICANT UNITS IN CRYAN'S BUCKMOTH | Biodiversity Conservation; Ecology; Environmental Sciences |
| 70 | STEVENS PF, 1991, SYST BOT, V16, P553 | 154,01 | CHARACTER STATES, MORPHOLOGICAL VARIATION, AND PHYLOGENETIC ANALYSIS - A REVIEW | Plant Sciences; Evolutionary Biology |

## CLUSTER 3

| # | Reference (WoS format) | Sum of weights | Title | WoS Category (Subject Category) |
|---|---|---|---|---|
| 1 | MILLS LS, 1993, BIOSCIENCE, V43, P219 | 1048,4 | THE KEYSTONE-SPECIES CONCEPT IN ECOLOGY AND CONSERVATION | Biology |

| # | Reference (WoS format) | Sum of weights | Title | WoS Category (Subject Category) |
|---|---|---|---|---|
| 2 | MENGE BA, 1994, ECOL MONOGR, V64, P249 | 1023,81 | THE KEYSTONE SPECIES CONCEPT - VARIATION IN INTERACTION STRENGTH IN A ROCKY INTERTIDAL HABITAT | Ecology |
| 3 | Power ME, 1996, BIOSCIENCE, V46, P609 | 633,95 | CHALLENGES IN THE QUEST FOR KEYSTONES | Biology |
| 4 | PAINE RT, 1992, NATURE, V355, P73 | 328,07 | FOOD-WEB ANALYSIS THROUGH FIELD MEASUREMENT OF PER-CAPITA INTERACTION STRENGTH | Multidisciplinary Sciences |
| 5 | WOOTTON JT, 1994, ECOLOGY, V75, P151 | 325,76 | PREDICTING DIRECT AND INDIRECT EFFECTS - AN INTEGRATED APPROACH USING EXPERIMENTS AND PATH-ANALYSIS | Ecology |
| 6 | WOOTTON JT, 1994, ANNU REV ECOL SYST, V25, P443 | 279,22 | THE NATURE AND CONSEQUENCES OF INDIRECT EFFECTS IN ECOLOGICAL COMMUNITIES | Ecology; Evolutionary Biology |
| 7 | WOOTTON JT, 1993, AM NAT, V141, P71 | 218,42 | INDIRECT EFFECTS AND HABITAT USE IN AN INTERTIDAL COMMUNITY - INTERACTION CHAINS AND INTERACTION MODIFICATIONS | Ecology; Evolutionary Biology |
| 8 | POWER ME, 1995, TRENDS ECOL EVOL, V10, P182 | 198,33 | THE KEYSTONE COPS MEET IN HILO | Ecology; Evolutionary Biology; Genetics & Heredity |
| 9 | TILMAN D, 1994, NATURE, V367, P363 | 186,15 | BIODIVERSITY AND STABILITY IN GRASSLANDS | Multidisciplinary Sciences |
| 10 | Wootton JT, 1997, ECOL MONOGR, V67, P45 | 168,88 | ESTIMATES AND TESTS OF PER CAPITA INTERACTION STRENGTH: DIET, ABUNDANCE, AND IMPACT OF INTERTIDALLY FORAGING BIRDS | Ecology |
| 11 | LAWTON JH, 1992, NATURE, V355, P19 | 164,03 | ECOLOGY - FEEBLE LINKS IN FOOD WEBS | Multidisciplinary Sciences |
| 12 | YODZIS P, 1988, ECOLOGY, V69, P508 | 159,42 | THE INDETERMINACY OF ECOLOGICAL INTERACTIONS AS PERCEIVED THROUGH PERTURBATION EXPERIMENTS | Ecology |
| 13 | Leibold MA, 1996, AM NAT, V147, P784 | 158,49 | A GRAPHICAL MODEL OF KEYSTONE PREDATORS IN FOOD WEBS: TROPHIC REGULATION OF ABUNDANCE, INCIDENCE, AND DIVERSITY PATTERNS IN COMMUNITIES | Ecology; Evolutionary Biology |
| 14 | COX PA, 1991, CONSERV BIOL, V5, P448 | 154,01 | FLYING FOXES AS STRONG INTERACTORS IN SOUTH-PACIFIC ISLAND ECOSYSTEMS - A CONSERVATION HYPOTHESIS | Biodiversity Conservation; Ecology; Environmental Sciences |

## CLUSTER 4

| # | Reference (WoS format) | Sum of weights | Title | WoS Category (Subject Category) |
|---|---|---|---|---|
| 1 | RIDLEY M, 1989, BIOL PHILOS, V4, P1 | 339,48 | THE CLADISTIC SOLUTION TO THE SPECIES PROBLEM | History & Philosophy Of Science |
| 2 | FROST DR, 1990, HERPETOLOGICA, V46, P87 | 289,21 | SPECIES IN CONCEPT AND PRACTICE - HERPETOLOGICAL APPLICATIONS | Zoology |

| #  | Reference (WoS format) | Sum of weights | Title | WoS Category (Subject Category) |
|----|------------------------|----------------|-------|----------------------------------|
| 3  | DEQUEIROZ K, 1988, CLADISTICS, V4, P317 | 191,3 | PHYLOGENETIC SYSTEMATICS AND THE SPECIES PROBLEM | Evolutionary Biology |
| 4  | HULL DL, 1976, SYST ZOOL, V25, P174 | 167,26 | ARE SPECIES REALLY INDIVIDUALS | Zoology |
| 5  | SIMONETTA AM, 1992, B ZOOL, V59, P447 | 164,03 | PROBLEMS OF SYSTEMATICS .1. A CRITICAL-EVALUATION OF THE SPECIES PROBLEM AND ITS SIGNIFICANCE IN EVOLUTIONARY BIOLOGY | Zoology |
| 6  | HULL DL, 1978, PHILOS SCI, V45, P335 | 154,52 | MATTER OF INDIVIDUALITY | History & Philosophy Of Science |
| 7  | KITCHER P, 1984, PHILOS SCI, V51, P308 | 149,04 | SPECIES | History & Philosophy Of Science |
| 8  | SIMONETTA AM, 1995, B ZOOL, V62, P37 | 148,74 | SOME REMARKS ON THE INFLUENCE OF HISTORICAL BIAS IN OUR APPROACH TO SYSTEMATICS AND THE SO-CALLED SPECIES PROBLEM | Zoology |
| 9  | WILEY EO, 1978, SYST ZOOL, V27, P17 | 137,35 | EVOLUTIONARY SPECIES CONCEPT RECONSIDERED | Zoology |
| 10 | SIMONETTA AM, 1993, B ZOOL, V60, P323 | 131,05 | PROBLEMS OF SYSTEMATICS .2. THEORY AND PRACTICE IN PHYLOGENETIC STUDIES AND IN SYSTEMATICS | Zoology |
| 11 | KLUGE AG, 1990, BIOL PHILOS, V5, P417 | 108,45 | SPECIES AS HISTORICAL INDIVIDUALS | History & Philosophy Of Science |
| 12 | NIXON KC, 1990, CLADISTICS, V6, P211 | 108,45 | AN AMPLIFICATION OF THE PHYLOGENETIC SPECIES CONCEPT | Evolutionary Biology |